\documentclass[12pt]{article}
\usepackage{amsmath}
\usepackage{amsfonts}
\usepackage{amssymb}
\usepackage{epsfig}

\makeatletter
\def\ifundefined{\@ifundefined}
\makeatother

\begin{document}

\title{Revealing Relationships among Relevant Climate Variables with Information Theory}


\author{\large Kevin H. Knuth{\normalsize\textsuperscript{1}},
 Anthony Gotera{\normalsize\textsuperscript{2,3}}, Charles T. Curry{\normalsize\textsuperscript{4,5}},\\
 Karen A. Huyser{\normalsize\textsuperscript{2,6}},
 Kevin R. Wheeler{\normalsize\textsuperscript{1}},\\
 William B. Rossow{\normalsize\textsuperscript{7}} \vspace{1ex}\\
 \normalsize \textsuperscript{1}Intelligent Sys. Div., NASA Ames Res. Center, Moffett Field CA, USA\\
 \normalsize \textsuperscript{2}Ed. Assoc., NASA Ames Res. Center, Moffett Field CA, USA\\
 \normalsize \textsuperscript{3}Dept. of Math., CSU East Bay, Hayward CA, USA\\
 \normalsize \textsuperscript{4}QSS, NASA Ames Res. Center, Moffett Field CA, USA\\
 \normalsize \textsuperscript{5}Dept. of App. Math. and Stat., UC Santa Cruz, Santa Cruz CA, USA\\
 \normalsize \textsuperscript{6}Dept. of Elect. Eng., Stanford Univ., Stanford CA, USA\\
 \normalsize \textsuperscript{7}NASA Goddard Inst. for Space Studies, New York NY, USA\\
}

\maketitle

\begin{abstract}
A primary objective of the NASA Earth-Sun Exploration Technology
Office is to understand the observed Earth climate variability,
thus enabling the determination and prediction of the climate's
response to both natural and human-induced forcing.  We are
currently developing a suite of computational tools that will
allow researchers to calculate, from data, a variety of
information-theoretic quantities such as mutual information, which
can be used to identify relationships among climate variables, and
transfer entropy, which indicates the possibility of causal
interactions.  Our tools estimate these quantities along with
their associated error bars, the latter of which is critical for
describing the degree of uncertainty in the estimates. This work
is based upon optimal binning techniques that we have developed
for piecewise-constant, histogram-style models of the underlying
density functions. Two useful side benefits have already been
discovered.  The first allows a researcher to determine whether
there exist sufficient data to estimate the underlying probability
density.  The second permits one to determine an acceptable degree
of round-off when compressing data for efficient transfer and
storage. We also demonstrate how mutual information and transfer
entropy can be applied so as to allow researchers not only to
identify relations among climate variables, but also to
characterize and quantify their possible causal interactions.
\end{abstract}


\section{Introduction}
A primary objective of the NASA Earth-Sun Exploration Technology
Office is to understand the observed Earth climate variability,
and determine and predict the climate's response to both natural
and human-induced forcing.  Central to this problem is the concept
of feedback and forcing.  The basic idea is that changes in one
climate subsystem will cause or force responses in other
subsystems.  These responses in turn feed back to force other
subsystems, and so on.  While it is commonly assumed that these
interactions can be described by linear systems techniques, one
must appeal to large-scale averages, asymptotic distributions and
central limit theorems to defend such models. In doing so, our
ability to describe processes with reasonably high spatiotemporal
resolution is lost in the averaging step.  There are distinct
advantages to developing feedback and forcing models that allow
for nonlinearity. This is especially highlighted by the results of
Lorenz's work in modelling convection cells \cite{lorenz}, which
is used today as a textbook example of a nonlinear system, and
historically was instrumental in the development of modern
nonlinear dynamics.

In the early stages of a field of science, much effort goes into
identifying the \emph{relevant variables}.  This is typically a
small set of variables that are used as parameters in idealized
scientific models of the physical phenomenon under study.  In
Galileo's time, he found that motion was best described by the
relevant variables: displacement, velocity, and acceleration.
Sometimes these scientific models are gross oversimplifications
that merely capture the basic essence of a physical process, and
sometimes they are highly detailed and allow one to make specific
predictions about the system.  In Earth Science, the fact that the
majority of our efforts are spent on amassing large amounts of
data indicates that we have not yet identified the relevant
variables for many of the problems that we study.  One of the aims
of this work is to develop methods that will enable us to better
identify relevant variables.

A second aim of this work is to develop techniques that will allow
us to identify relationships among these relevant variables.  As
mentioned above, it is naive to expect that these variables will
interact linearly.  Thus techniques that are sensitive to both
linear and nonlinear relationships will better enable us to
identify interactions among these variables. Information theory
allows one to compute the amount of information that knowledge of
one variable provides about another
\cite{Shannon&Weaver,Cover&Thomas}. Such computations are
applicable to both linear and nonlinear relationships between the
variables.  Furthermore, they rely on higher-order statistics;
whereas approaches such as correlation analysis, Empirical
Orthogonal Functions (EOF), Principal Component Analysis (PCA),
and Granger causality \cite{Granger} are based on second-order
statistics, which amount to approximating everything with Gaussian
distributions. An additional benefit is the fact that higher order
generalizations of basic information-theoretic quantities are
deeply connected to the concept of relevance
\cite{Knuth:MaxEnt2004,Knuth:2005}, and thus this approach is the
natural methodology for identifying relevant variables and their
interactions with one another.

Information-theoretic computations ultimately rely on quantities
such as entropy.  While researchers have been estimating entropy
from data for years, relatively few attempts have been made to
estimate the uncertainties associated with the entropy estimates.
We consider this to be of paramount importance, since the degree
to which we understand the Earth's climate system can only be
characterized by quantifying our uncertainties.  The remainder of
this paper will describe our ongoing efforts to estimate
information-theoretic quantities from data as well as the
associated uncertainties, and to demonstrate how these approaches
will be used to identify relationships among relevant climate
variables.

\section{Density Models}
Our knowledge about a variable depends on what we know about the
values that it can take.  For instance, knowing that the average
daytime summer beach water temperature in Hawaii is $80^\circ$F
provides some information.  However, more information would be
provided by the variance of this quantity.  A complete
quantification of our knowledge of this variable would be given by
the probability density function.  From that, one can compute the
probability that the water temperature will fall within a given
range.  To apply these information-theoretic techniques, we first
must estimate the probability density function from a data set.

\begin{figure}[t]
\begin{center}
\includegraphics[width=0.6\textwidth]{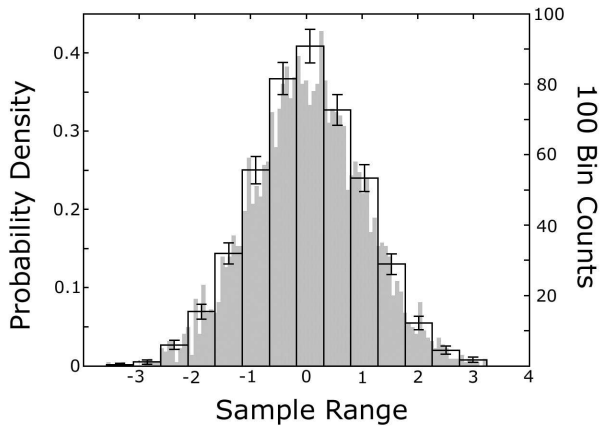}
\end{center}
\caption{\label{fig:pdfmodel}The optimal piecewise-constant
probability density model generated from 1000 data samples drawn
from a Gaussian density. The error bars indicate the uncertainty
in the bin heights. It is superimposed over a 100-bin histogram
that shows the irrelevant sampling variations of the data.}
\end{figure}

\subsection{Piecewise-Constant Density Models}
We model the density function with a piecewise-constant model.
Such a model divides the range of values of the variable into a
set of $M$ discrete bins and assigns a probability to each bin. We
denote the probability that a data point is found to be in the
$k^{th}$ bin by $\pi_k$. The result is closely related to a
histogram, except that the ``height'' of the bin $h_k$, is the
constant probability density (bin probability divided by the bin
width) over the region of the bin. Integrating this constant
probability density $h_k$ over the width of the bin $v_k$ leads to
a total probability $\pi_k = h_k v_k$ for the bin. This leads to
the following piecewise-constant model $h(x)$ of the unknown
probability density function for the variable $x$
\begin{equation}
h(x) = \sum_{k = 1}^{M}{h_k~\Pi(x_{k-1}, x, x_k)},
\end{equation}
where $h_k$ is the probability density of the $k^{th}$ bin with
edges defined by $x_{k-1}$ and $x_k$, and $\Pi(x_{k-1}, x, x_k)$
is the boxcar function where
\begin{equation}
\Pi(x_a, x, x_b) =
    \left\{ \begin{array}{rl}
       0 & \mbox{if}~~x < x_a\\
       1 & \mbox{if}~~x_a \leq x < x_b\\
       0 & \mbox{if}~~x_b \leq x \end{array} \right.
\end{equation}
For the case of equal bin widths, this density model can be
re-written in terms of the bin probabilities $\pi_k$ as
\begin{equation}
h(x) = \frac{M}{V} \sum_{k = 1}^{M}{\pi_k~\Pi(x_{k-1}, x, x_k)}.
\end{equation}
where $V$ is the width of the entire region covered by the density
model.  This formalism is readily expanded into multiple
dimensions by extending $k$ to the status of a multi-dimensional
index, and using $v_k$ to represent the multi-dimensional volume,
with $V$ representing the multi-dimensional volume covered by the
density model.

To accurately describe the density function, we use the data to
compute the optimal number of bins.  This is performed by applying
Bayesian probability theory \cite{Sivia:1996,Gelman+etal:1995} and
writing the posterior probability of the model parameters
\cite{Knuth:optBINS}, which are the number of bins $M$ and the bin
probabilities $\underline{\pi} = \{\pi_1, \pi_2, \ldots,
\pi_{M-1}\}$,\footnote{Note that there are only $M-1$ bin
probability parameters since there is a constraint that the sum
should add to one.} as a function of the $N$ data points
$\underline{d} = \{d_1, d_2, \ldots, d_N\}$
\begin{align} \label{eq:joint-posterior}
p(\underline{\pi}, M | \underline{d}, I)& \propto
\biggl(\frac{M}{V}\biggr)^N
\frac{\Gamma\bigl(\frac{M}{2}\bigr)}{\Gamma\bigl(\frac{1}{2}\bigr)^M}~~~\times\\
& \pi_1^{n_1-\frac{1}{2}} \pi_2^{n_2-\frac{1}{2}} \ldots
\pi_{M-1}^{n_{M-1}-\frac{1}{2}}
\biggl(1-\sum_{i=1}^{M-1}{\pi_i}\biggr)^{n_M-\frac{1}{2}},\nonumber
\end{align}
where $n_k$ is the number of data points in the $k^{th}$ bin. Note
that the symbol $I$ is used to represent any prior information
that we may have or any assumptions that we have made, such as
assuming that the bins are of equal width.

Integrating over all possible bin heights gives the marginal
posterior probability of the number of bins given the data
\cite{Knuth:optBINS}
\begin{equation} \label{eq:posterior-for-M}
p(M | \underline{d}, I) \propto \biggl(\frac{M}{V}\biggr)^N
\frac{\Gamma\bigl(\frac{M}{2}\bigr)}{\Gamma\bigl(\frac{1}{2}\bigr)^M}~
\frac{\prod_{k=1}^{M}{\Gamma(n_k+\frac{1}{2})}}{\Gamma(N+\frac{M}{2})},
\end{equation}
where the $\Gamma(\cdot)$ is the Gamma function
\cite[p.~255]{Abramowitz&Stegun}. The idea is to evaluate this
posterior probability for all the values of the number of bins
within a reasonable range and select the result with the greatest
probability. In practice, it is often much easier computationally
to search for the optimal number of bins $M$ by finding the value
of $M$ that maximizes the logarithm of the probability,
(\ref{eq:posterior-for-M}) above.

Using the joint posterior probability (\ref{eq:joint-posterior})
one can compute the mean bin probabilities and the standard
deviations from the data \cite{Knuth:optBINS}.  The mean bin
probability is
\begin{equation} \label{eq:mean-prob}
\mu_k = \langle h_k \rangle = \frac{\langle \pi_k \rangle}{v_k} =
\biggl(\frac{M}{V}\biggr)
\biggl(\frac{n_k+\frac{1}{2}}{N+\frac{M}{2}}\biggr),
\end{equation}
and the associated variance of the height of the $k^{th}$ bin is
\begin{equation} \label{eq:var-prob}
\sigma_k^2 = \biggl(\frac{M}{V}\biggr)^2
\biggl(\frac{(n_k+\frac{1}{2})(N-n_k+\frac{M-1}{2})}
{(N+\frac{M}{2}+1)(N+\frac{M}{2})^2}\biggr),
\end{equation}
where the standard deviation is the square root of the variance.
Note that bins with no counts still have a non-zero probability.
No lack of evidence can ever prove conclusively that an event
occurring in a given bin is impossible---just less probable.

In this way we are able to estimate probability densities from
data, and quantify the uncertainty in our knowledge. An example of
a probability density model is shown in Figure \ref{fig:pdfmodel}.
This optimal binning technique ensures that our density model
includes all the relevant information provided by the data while
ignoring irrelevant details due to sampling variations.  The
result is the most honest summary of our knowledge about the
density function from the given data. Honest representations are
important since they can reveal two potentially disastrous
situations: insufficient data and excessive round-off error.

\subsection{Insufficient Data}
Without examining the uncertainties, one can never be sure that
one has a sufficient amount of data to make an inference.  How
many data points does one need to estimate a density function?  Do
we need $100$ data points? $10000$? a million?

\begin{figure}
\begin{center}
\includegraphics[width=0.70\textwidth]{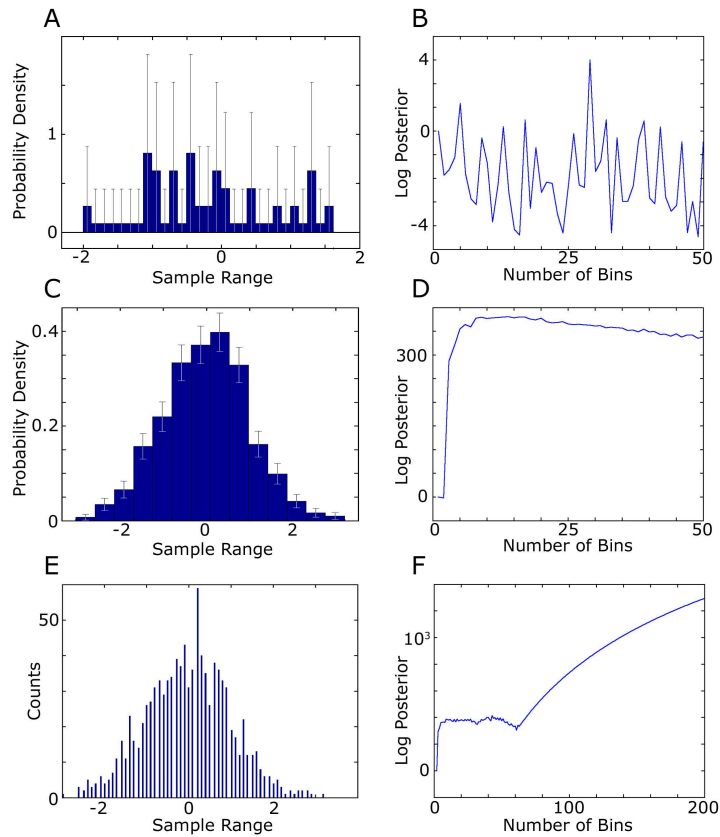}
\end{center}
\caption{\label{fig:examples} A) With a small number of data
points (in this case $30$ data points sampled from a Gaussian), it
is not possible to determine the probability density with any
accuracy. B) The log posterior in this case jumps around with many
spurious local maxima.  This indicates that the inference is
unreliable and more data are needed. C) With a sufficient amount
of data (in this case $1000$ data points), the probability density
is easily estimated. D) The log posterior rises rapidly as the
number of bins $M$ increases reaching a peak (in this case at
$M=14$) and then falling off gently. E) In this example, we take
the previous 1000 data points and rounded them off to the nearest
$1/10$th.  As a result of this severe truncation, the optimal
solution looks like a picket fence.  The discrete nature of the
truncated data is now a more prominent feature than the original
probability density function. F) The log posterior detects this
once the data have been separated into the discrete bins and can
be separated no further.}
\end{figure}

By examining the log posterior probability for the optimal number
of bins given the data, one can easily detect whether one
possesses sufficient data.\footnote{The uncertainties, error bars
or standard deviations are summary quantities that characterize
the behavior of the posterior probability around the optimal
solution. For this reason, rather than computing uncertainties, we
simply look at the log posterior.}  In this example (Figure
\ref{fig:examples}), we see two density models constructed from
data sampled from a Gaussian distribution. In the first case, we
have collected $30$ data points, and in the second case, $1000$.
In the case of $30$ data points, the behavior of the log posterior
probability, which is the logarithm of (\ref{eq:posterior-for-M}),
is very noisy with spurious maxima. We can not be sure how many
bins to use, and are thus very uncertain as to the shape of the
density function from which these data were sampled. In the case
with $1000$ data points, the behavior of the log posterior is
clear. It rises sharply as the number of proposed bins increases
and reaches a peak and then gently falls off.  The result is an
estimate of the number of bins that provides a piecewise-constant
density model that, given the data, optimally describes the true
``unknown'' Gaussian distribution.

In our numerical experiments, we have found that for Gaussian
distributed data, one needs approximately $75$ to $100$ data
points to get a reasonable solution, and $150$ to $200$ data
points to be very certain. Note that this is a different situation
than assuming that we know that the underlying distribution is
Gaussian and then trying to estimate the mean and variance.  That
is a very different problem where the prior knowledge that it is
Gaussian (which would be represented by that $I$ again) makes is
feasible to make the inference using significantly fewer data
points.

\subsection{Excessive Round-Off}
The other problem that can occur is loss of information due to
data compression or round-off.  Many times to save memory space,
data values are truncated to a small number of decimal places.
When it is not clear how much information the data contains, it is
not clear to what degree the data can be truncated before
destroying valuable information.  Our optimal binning technique is
useful here as well.

In the event that the data has been severely truncated, the
optimal binning algorithm will see the discrete structure in the
data due as being more meaningful than the overall shape of the
underlying density function (Figure \ref{fig:examples}E). The
result is that the optimal number of bins leads to what is called
``picket fencing'', where the density model looks like a picket
fence. There is no graceful way to recover from this---relevant
data has been lost, and cannot be recovered.

\section{Entropy and Information}
We can characterize the behavior of a system $X$ by looking at the
set of states the system visits as it evolves in time.  If a state
is visited rarely, we would be surprised to find the system there.
We can express the expectation (or lack of expectation) to find
the system in state $x$ in terms of the probability that it can be
found in that state, $p(x)$, by
\begin{equation}
s(x) = \log \frac{1}{p(x)}.
\end{equation}
This quantity is often called the surprise, since it is large for
improbable events and small for probable ones.  Averaging this
quantity over all of the possible states of the system gives a
measure of our expectation of the state of the system
\begin{equation}
H(X) = \sum_{x \in X}{p(x) \log \frac{1}{p(x)}}.
\end{equation}
This quantity is called the \emph{Shannon Entropy}, or entropy for
short \cite{Shannon&Weaver}.  It can be thought of as a measure of
the amount of information we possess about the system.  It is
usually expressed by rewriting the fraction above using the
properties of the logarithm
\begin{equation}\label{eq:entropy}
H(X) = -\sum_{x \in X}{p(x) \log {p(x)}}.
\end{equation}
Note that changing the base of the logarithm merely changes the
units in which entropy is measured.  When the logarithm base is 2,
entropy is measured in \emph{bits}, and when it is base $e$, it is
measured in \emph{nats}.

If the system states can be described with multiple parameters, we
can use them jointly to describe the state of the system. The
entropy can still be computed by averaging over all possible
states.  For two subsystems $X$ and $Y$ the \emph{joint entropy}
is
\begin{equation}\label{eq:joint-entropy}
H(X,Y) = -\sum_{x \in X}{\sum_{y \in Y}{p(x,y) \log {p(x,y)}}}.
\end{equation}

The differences of entropies are useful quantities.  Consider the
difference between the joint entropy $H(X,Y)$ and the individual
entropies $H(X)$ and $H(Y)$
\begin{equation}\label{eq:mi}
MI(X,Y) = H(X) + H(Y) - H(X,Y).
\end{equation}
This quantity describes the difference in the amount of
information one possesses when one considers the system jointly
instead of considering the system as two individual subsystems. It
is called the \emph{Mutual Information} (MI) since it describes
the amount of information that is shared between the two
subsystems. If you know something about subsystem $X$, the mutual
information describes how much information you also possess about
$Y$, and vice versa.  Thus MI quantifies the relevance of
knowledge about one subsystem to knowledge about another
subsystem.  For this reason, it is useful for identifying and
selecting a set of relevant variables that can aid in the
prediction of another climate variable. One should note that if
two climate variables X and Y are independent, then $H(X, Y) =
H(X) + H(Y)$, then the mutual information (\ref{eq:mi}) is
zero---as one would expect.  The mutual information is a measure
of true statistical independence, whereas concepts like
decorrelation only describe independence up to second-order. Two
variables can be uncorrelated, yet still dependent.\footnote{This
fact is usually poorly understood and it stems from the confusion
between the common meaning of the word `uncorrelated', which we
usually take to mean ``independent'', and the precise mathematical
definition of the word ``uncorrelated'', which means that the
covariance matrix is of diagonal form.}

While the mutual information is an important quantity in
identifying relationships between system variables, it provides no
information regarding the causality of their interactions.  The
easiest way to see this is to note that the mutual information is
symmetric with respect to interchange of X and Y, whereas causal
interactions are not symmetric.  To identify causal interactions,
a asymmetric quantity must be utilized.  Recently, Schreiber
\cite{Schreiber:2000} introduced a novel information-theoretic
quantity called the \emph{Transfer Entropy} (TE).  Consider two
subsystems $X$ and $Y$, with data in the form of two time series
of measurements
\begin{eqnarray}
X &=& \{x_1, x_2, \ldots, x_t, x_{t+1}, \ldots, x_n\}\nonumber \\
Y &=& \{y_1, y_2, \ldots, y_s, y_{s+1}, \ldots, y_n\}\nonumber
\end{eqnarray}
with $t=s+l$ where $l$ is some lag time.  The transfer entropy can
be written as
\begin{equation}
T(X_{t+1}|X_t, Y_s) = I(X_{t+1},Y_s) - I(X_t, X_{t+1}, Y_s)
\end{equation}
where $I(X_{t+1},Y_s)$ is the rank-2 co-information (mutual
information) and $I(X_t, X_{t+1}, Y_s)$ is the rank-3
co-information, which describes the information that all three
variables share \cite{Bell:2003,Knuth:2005}.  Thus the transfer
entropy is just the information shared by Y and future values of X
minus the information shared by Y , X, and future values of X.  In
this way it captures the predictive information Y possesses about
X and thus is an indicator of a possible causal interaction. Using
the definitions of these higher-order informations, the TE can be
re-written in the more convenient, albeit less intuitive form,
originally suggested by Schreiber \cite{Schreiber:2000}
\begin{align}
&T(X_{t+1}|X_t, Y_s) = \\
&-H(X_t) + H(X_t, Y_s) + H(X_t, X_{t+1}) - H(X_t, X_{t+1},
Y_s),\nonumber
\end{align}
where $H(X_t, X_{t+1}, Y_s)$ is the joint entropy between the
subsystems $X$, $Y$, and a time-shifted version of $X$, $X_{t+1}$.
Unlike the mutual information, TE is not symmetric with
interchange of $X$ and $Y$
\begin{align}
&T(Y_{s+1}|X_t, Y_s) = \\
&-H(Y_s) + H(X_t, Y_s) + H(Y_s, Y_{s+1}) - H(X_t, Y_s,
Y_{s+1}).\nonumber
\end{align}
This asymmetry is crucial since it is indicative of the ability of
TE to identify causal interactions.

This is the basic outline of the theory, the next section deals
with the practical considerations of estimating these quantities
from data and obtaining error bars to indicate the uncertainties
in our estimates.

\section{Estimating Entropy and Information}
Given a multi-dimensional data set, we begin by estimating the
number of bins that will provide an optimal probability density
model.  With this probability density model in hand, we can begin
computing the information-theoretic quantities described above.
The challenge is to propagate our uncertainties in our knowledge
about the probability density to uncertainties in our knowledge
about the entropy, mutual information, and transfer entropy
estimates.

\begin{figure}[t]
\begin{center}
\includegraphics[width=0.7\textwidth]{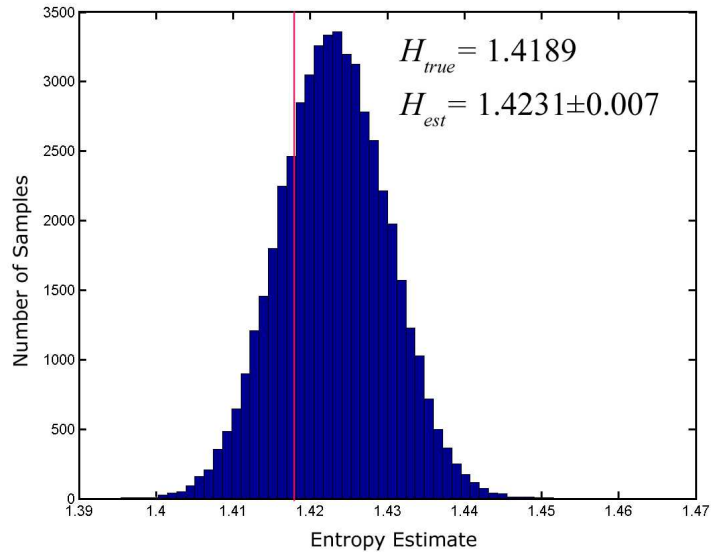}
\end{center}
\caption{\label{fig:entropy}This figure shows the histogram of the
entropy samples computed from the bin probabilities drawn from a
Dirichlet distribution defined by the data.  The true entropy
falls within one standard deviation of our estimate.}
\end{figure}

Say that you have a variable that you know is Gaussian distributed
with zero mean and unit variance, $\mathcal{N}(0,1)$.  If you want
to obtain an instance of this variable that is in accordance with
its known Gaussian probability density, you merely need to sample
a point from a Gaussian distribution with zero mean and unit
variance.  It is easy to obtain many such instances by generating
many samples, and it is not surprising to find that the mean and
variance of those instances is consistent with the density from
which they were sampled.

We take the same approach here. Given the number of bins $M$ in
the probability density model, the posterior probability
(\ref{eq:joint-posterior}) of the bin heights has the form of a
Dirichlet distribution.  One can sample the bin heights from the
Dirichlet distribution by sampling each bin height from a gamma
distribution with common scale and shape parameters and
renormalizing the resulting set to unit probability
\cite[p.~482]{Gelman+etal:1995}. Every set of bin height samples
that is drawn, constitutes a probability density model that could
very well describe the given data.  By taking something on the
order of $50,000$ samples, we have a set of $50,000$ probability
density models each of which are probable descriptions of the
data.  The fact that we get many different, albeit similar,
density models is a result of the fact that we are uncertain as to
which model is correct.  Without an infinite amount of data, we
will always be uncertain---the question is: how uncertain?  By
simply computing the mean and variance of the bin heights from
this set of samples, we can confirm that it approaches the
theoretical mean (\ref{eq:mean-prob}), and likewise with the
height variance (\ref{eq:var-prob}). This sample variance, or its
square root---the standard deviation, of the bin heights
quantifies our uncertainty about the probability density.

For each sampled probability density model, we can compute the
entropy.  This will be given by (\ref{eq:entropy}) for a
one-dimensional density function, by (\ref{eq:joint-entropy}) for
a two-dimensional density function, and so on for higher
dimensions.  The result is a list of $50,000$ or so entropies,
from which we can readily compute the mean and standard deviation
thus providing us with an entropy estimate and an associated
standard deviation quantifying our uncertainty.

In one experiment, $10,000$ data points were sampled from a
Gaussian distribution with zero mean and unit variance. The
optimal number of bins was found to be $M=24$.  The number of
counts per bin for each of the $24$ bins was used to sample
$50,000$ probability density models from a Dirichlet distribution.
From each of these samples, the entropy was computed. Figure
\ref{fig:entropy} shows a histogram of the $50,000$ entropy
samples. The mean entropy was found to be $H_{est} =
1.4231\pm0.007$. The true entropy, which is $H_{true} = 1.1489$,
is within one standard deviation of our estimate. This indicates
that $H_{est}$ is a reasonable estimate of the entropy that
simultaneously quantifies our uncertainty as to its precise value.

The mutual information and transfer entropy are computed
similarly, with the understanding that to compute the mutual
information, one works with two-dimensional density functions, and
for the transfer entropy one works with three-dimensional
densities.  Despite the increase in dimensionality, the sampling
procedure works exactly as described above for the one-dimensional
case.

\begin{figure}[t]
\begin{center}
\includegraphics[width=0.80\textwidth]{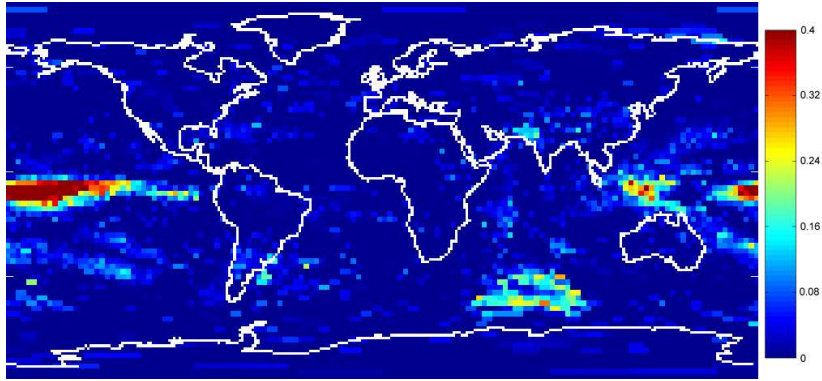}
\end{center}
\caption{\label{fig:cloud-sst}This figure shows a preliminary
mutual information map, which quantifies the relationship between
the Cold Tongue Index, which is indicative of the equatorial
Pacific sea surface temperatures, and the Percent Cloud Cover
across the globe.}
\end{figure}

\section{Application to Climate Variables}
To demonstrated that the mutual information can identify
relationships between climate variables, we performed several
preliminary explorations.  In one of our explorations, we
considered the percent cloud cover (computed as a monthly average)
as one subsystem $X$. These data were obtained from the
International Satellite Cloud Climatology Project (ISCCP) climate
summary product C2
\cite{Schiffer&Rossow:1983,Rossow&Schiffer:1999}, and consisted of
monthly averages of percent cloud cover resulting in a time-series
of 198 months of 6596 equal-area pixels each with side length of
280 km.  It is best to think of the percent cloud cover at each
pixel as an independent subsystem, say $X_1, X_2, \ldots,
X_{6596}$. The other subsystem $Y$ was chosen to be the Cold
Tongue Index (CTI), which describes the sea surface temperature
anomalies in the eastern equatorial Pacific Ocean (6N-6S,
180-90W)\cite{Deser&Wallace:1990}. These anomalies are known to be
indicative of the El Ni\~{n}o Southern Oscillation
(ENSO)\cite{Deser&Wallace:1987,Mitchell&Wallace:1996}. Thus the
second subsystem $Y$ consists of the set of 198 monthly values of
CTI, and corresponds in time to the cloud cover subsystems.

The mutual information was computed between $X_1$ and $Y$, and
$X_2$ and $Y$, and so on by using (\ref{eq:mi}). This enables us
to generate a global map of 6596 mutual information calculations
(Figure \ref{fig:cloud-sst}), which indicates the relationship
between the Cold Tongue Index (CTI) and percent cloud cover across
the globe. Note that the cloud cover affected by the sea surface
temperature (SST) variations lies mainly in the equatorial
Pacific, along with an isolated area in Indonesia. The highlighted
areas in the Indian longitudes are known artifacts of satellite
coverage.

\begin{figure}[t]
\begin{center}
\includegraphics[width=0.6\textwidth]{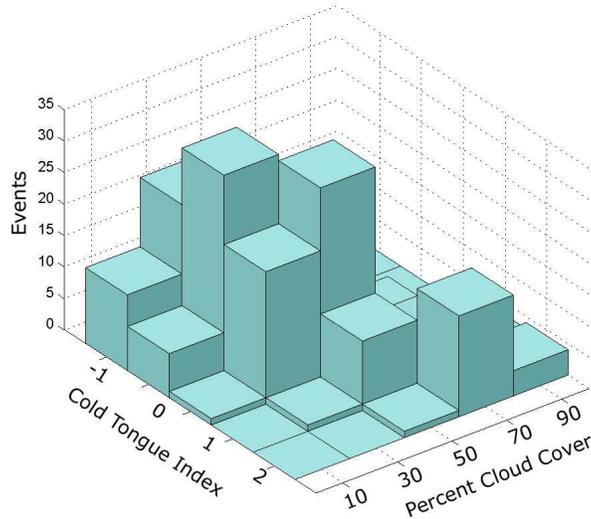}
\end{center}
\caption{\label{fig:cloud-sst-histo}This figure shows the optimal
histogram (error bars omitted) of the joint data formed by
combining the Cold Tongue Index time series with the Percent Cloud
Cover time series at the location where the mutual information was
found to be maximal.}
\end{figure}

Pixel $3231$, which lies in the equatorial Pacific (1.25N
191.25W), was found to have the greatest mutual information. Thus
cloud cover at this point is maximally relevant to the CTI and
vice versa.  By taking the time series representing the percent
cloud cover at this position, we can combine this with the CTI
time series to construct an optimal two-dimensional density model
(Figure \ref{fig:cloud-sst-histo}). This density function is not
factorable into the product of two independent one-dimensional
density functions.  This indicates that the mutual information is
non-zero (as we had previously determined), and that the two
quantities are related in the sense that one variable provides
information about the other.

We are currently working to sample these density functions from
their corresponding Dirichlet distributions to obtain more
accurate estimates of these information-theoretic quantities along
with error bars indicating the uncertainty in our estimates.  The
end result will be a set of software tools that will allow
researchers to rapidly and accurately estimate
information-theoretic measures to identify, qualify and quantify
causal interactions among climate variables from large climate
data sets.

\vspace{1ex}\hrule\vspace{3ex}

\end{document}